%% file: main.tex
\documentclass[iicol,sn-mathphys-num]{sn-jnl}

\usepackage{graphicx}%
\usepackage{multirow}%
\usepackage{amsmath,amssymb,amsfonts}%
\usepackage{amsthm}%
\usepackage{mathrsfs}%
\usepackage[title]{appendix}%
\usepackage{xcolor}%
\usepackage{textcomp}%
\usepackage{manyfoot}%
\usepackage{booktabs}%
\usepackage{algorithm}%
\usepackage{algorithmicx}%
\usepackage{algpseudocode}%
\usepackage{listings}%

\unnumbered %unnumbered Sections/subsections 

\begin{document}

\title{Supersolidity in Optically Trapped Polariton Condensates}

\author[1]{\fnm{P.\,N.} \sur{Kozhevin}}\email{pashakozhevin@gmail.com}

\author[2]{\fnm{A.\,D.} \sur{Liubomirov}}\email{lyubomirov\_ad@mail.ru}

\author[1,2]{\fnm{R.\,V.} \sur{Cherbunin}}\email{r.cherbunin@gmail.com}

\author[1]{\fnm{M.\,A.} \sur{Chukeev}}\email{maxchukeev@gmail.com}

\author[3]{\fnm{I.\,Yu.} \sur{Chestnov}}\email{igor\_chestnov@mail.ru}

\author[1,2,4]{\fnm{A.\,V.} \sur{Kavokin}}\email{kavokinalexey@gmail.com}

\author*[2,4]{\fnm{A.\,V.} \sur{Nalitov}}\email{nalitov.av@mipt.ru}

\affil[1]{\orgdiv{Department of Physics}, \orgname{St. Petersburg State University}, \orgaddress{\street{University Embankment, 7/9}, \city{St. Petersburg}, \postcode{199034}, \country{Russia}}}

\affil[2]{\orgname{Russian Quantum Center}, \orgaddress{\city{Skolkovo, Moscow}, \postcode{121205} \country{Russia}}}

\affil[3]{\orgname{School of Physics and Engineering, ITMO University}, \orgaddress{\street{Kronverksky Pr. 49, bldg. A}, \city{St. Petersburg}, \postcode{197101} \country{Russia}}}

\affil[4]{\orgname{Abrikosov Center for Theoretical Physics, MIPT}, \orgaddress{\street{Institutskiy per., 9}, \city{Dolgoprudny}, \state{Moscow Region}, \postcode{141701}, \country{Russia}}}

\abstract{
Superfluids under specific conditions can exhibit spontaneous breaking of continuous translation symmetries and form exotic spatially ordered states of matter known as supersolids.
Despite its early theoretical prediction, it took over half-a-centrury to experimentally demonstrate the supersolid phase in ultracold atomic Bose-Einstein condensates, forming due to long-range interatomic interactions.
Here we propose as a promising new platform for supersolidity exciton-polariton superfluids, confined in annular optically induced traps.
The supersolid phase emerges due to effective attractive interactions, mediated by the normal excitonic component of the system.
Experimental demonstration of spontaneously formed spatially ordered phase is in agreement with detailed mean-field theoretical analysis and numerical simulation.
The spontaneous character of the observed supersolid transition is further evidenced by the formation of specific zero-energy Nambu-Goldstone modes in the collective excitation spectrum.
}
\maketitle

The supersolid state of matter comprises properties of superfluidity, such as the absence of viscosity and coherent wave dynamics of matter at the macroscopic level, with long-range spatial ordering, characteristic of the solid state \cite{Chester1970,Leggett1970}.
Signatures of supersolidity were originally found in solid Helium-4 \cite{Kim2004}, but were later attributed to dislocation and Helium-3 impurities \cite{Day2007}.
Realization of supersolidity in ultra-cold atomic gases forming Bose-Einstein condensates was achieved later with the use of optical trapping and optical lattices \cite{Li2017,Lonard2017}.
The transition to the supersolid phase is accompanied by the spontaneous breaking of continuous translation symmetries, which was evidenced by the emergence of Nambu-Goldstone (N-G) modes in further experimental implementations of atomic supersolidity in the absence of an optical lattice \cite{Tanzi2019,Guo2019}.

Effective attraction is the key ingredient of the transition transition from the superfluid to supersolid phase in the rotonlike instability scenario \cite{Santos2003,Cinti2010,Henkel2010}.
Such partial attraction can be induced for artificial dipolar composite bosons in low-dimensional semiconductor structures \cite{Matuszewski2012}.
In driven-dissipative nonequilibrium bosonic condensates of exciton-polaritons, light-matter quasi-particles arising from strong coupling of excitonic and photonic modes \cite{KavokinMicrocavities}, the effective attraction can be mediated by the presence of an incoherent excitonic reservoir \cite{Vishnevsky2014}.
The reservoir-mediated effective attraction leads to spontaneous breaking of spatial symmetry and condensate self-trapping via the hole-burning effect \cite{Estrecho2018}.

At the same time, out-of-equilibrium polariton condensates exhibit superfluidity due to a combination of contact repulsion and extremely low effective masses \cite{Amo2011,Carusotto2013}.
The low effective masses of exciton-polaritons are determined by their photonic components and are responsible for room-temperature Bose-Einstein condensation.
%, inherited from exciton Coulomb exchange interaction, and extremely low polariton effective masses, stemming from the photonic part of polariton states and allowing high-temperature bosonic condensation of exciton-polaritons \cite{Amo2011,Carusotto2013}.
Weak repulsive interactions, governed by the exciton-exciton exhcange Coulomb interaction, result in linearization of the energy dispersion of condensate Bogoliubov excitations, which is a signature of superfluidity according to the Landau criterion \cite{Utsunomiya2008,Kohnle2011,Estrecho2021}.
Quantized vortices, vividly manifesting polariton superfluidity \cite{Lagoudakis2008} and responsible for the the Berezinski-Kosterlitz-Thouless transition \cite{Caputo2018}, were predicted to form spatially ordered rotating vortex lattices \cite{Keeling2008}, exhibiting similarity with lattices of Abrikosov vortices.

Spatially ordered polaritonic lattices can be produced explicitly by local modification of microcavity properties or by spatial profiling of optical pumping \cite{Schneider2017}.
However, spontaneous spatial ordering through the break of a discrete one-dimensional translation symmetry was first experimentally discovered in periodic polariton microwires \cite{Zhang2015}.
Recently, the formation of a density wave due to parametric scattering, attributed to supersolidity, was demonstrated in a photonic lattice supporting bound-in-the-continuum states \cite{Nigro2025,Trypogeorgos2025}.
A similar density wave due to ground-state degeneracy in spin-orbit coupled spinor polariton condensates was shown in liquid crystal microcavities \cite{Muszynski2024arxiv}.

\begin{figure}
    \includegraphics[width=\linewidth]{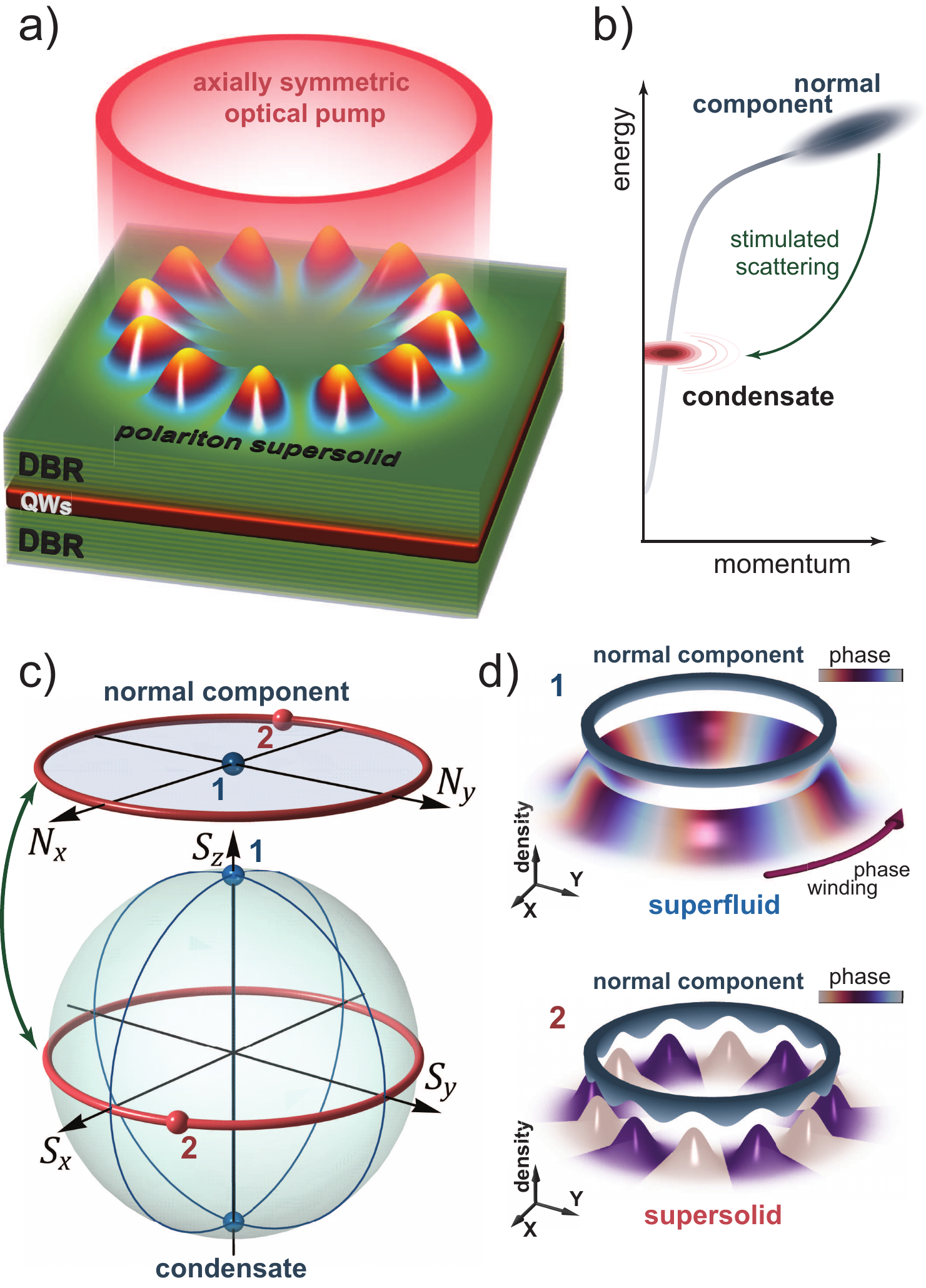}
    \caption{
    Sketch of the formation mechanism for polariton supersolids in optically induced traps.
    \textbf{a}, Real and \textbf{b}, reciprocal space pictures: incoherent spatially profiled optical pumping (red) generates an excitonic reservoir, forming a normal component (dark blue), from which the supersolid polaritonic condensate phase emerges (dark red).
    \textbf{c} Two-mode representation on the Poincar\'e sphere: two superfluid states at the poles and a ring of neutrally stable supersolid states on the equator.
    \textbf{d} Real space representations of the superfluid (winding phase) and supersolid (staggered phase) condensates with the respective normal components.}
    \label{Fig1}
\end{figure}  

In this work, we report the observation of polariton supersolidity in annular optically induced traps due to an interplay of an effective attractive polariton-polariton interaction mediated by the incoherent excitonic reservoir and the nonequilibrium nature of a polariton gas favouring condensation at high angular momentum states.
Bose-Einstein condensation of polaritons in this configuration was observed both in the thermodynamic equilibrium \cite{sun2017} and in the out-of-equilibrium regime \cite{Askitopoulos2013}.
Spontaneous breaking of the continuous time-translation symmetry and formation of dynamical phases similar to time crystallization was predicted in optically trapped polariton condensates \cite{Nalitov2019}.
Formation of petal-shaped condensates with orientation pinned by the sample disorder or trap asymmetry, was experimentally observed in annular optical traps \cite{Dreismann2014b,Sun2018}.
Our detailed experimental study of the extended spatial coherence and ordering of polariton condensates in axially symmetric traps demonstrates global coherence and crystalline order.
In addition, we relate the supersolid transition to the recently proposed instability mechanism of high-angular-momentum condensates due to the non-adiabatic character of the normal-to-superfluid coupling \cite{Chestnov2024}.
Finally, we note that the transition to the supersolid phase is accompanied with the spontaneous breaking of the continuous radial symmetry, which is evidenced in the emergence of the N-G mode in the collective excitation spectrum.

\section{Supersolid phase formation}

We create optically trapped polariton condensates in the nonresonant optical pumping scheme, where an isotropic annular profiled excitation light beam produced with a spatial light modulator, is focused onto the semiconductor planar microcavity as shown in Fig. \ref{Fig1}a (see Methods for details).
In the stationary regime, the axially symmetric optical pump induces an annular incoherent noncondensed excitonic gas.
It plays a role of the normal component of polariton fluid.
This component is formed by quasi-particles in the vicinity of the inflection point of the lower polariton dispersion branch, schematically shown in Fig. \ref{Fig1}b.
The incoherent annular reservoir simultaneously provides both trapping potential and gain for the nonequilibrium polariton condensate, thus suppressing losses related to the in-plane ballistic expansion and compensating the out-of-plane radiative losses.

We describe the optically trapped polariton supersolid with a mean-field model, based on the coupled equations for the superfluid and normal components given by the wave function $\Psi$ and the density $n$, respectively \cite{Wouters2007a,Haug2014}:
\begin{subequations}\label{eq:FullModel}
\begin{align}
i\hbar \partial_t \Psi &= \left[ - \frac{\hbar^2 \nabla^2}{2m} 
+ n \frac{\alpha + i \beta}{2} - i \hbar \frac{\Gamma}{2} + \alpha_1 |\Psi|^2 \right]\Psi, \label{eq:GPE} \\
\partial_t n &= P - \left( \gamma + \frac{\beta |\Psi|^2}{\hbar} \right)n. \label{eq:reservoir}
\end{align}
\end{subequations}
Here $m>0$ is the polariton effective mass, $\boldsymbol{\nabla}$ stands for the two-dimensional gradient operator, $\alpha>0$ is the strength of repulsive interaction between the polariton superfluid and the excitonic normal component, $\beta>0$ govern the rate of bosonic stimulated scattering from the normal excitons to the superfluid, $\alpha_1$ is the strength of the superfluid self-interaction, $\Gamma>0$ and $\gamma>0$ are the dissipation rates of the superfluid and the normal component.

The emergence of supersolidity in annular traps can be essentially described analytically within the two-mode approximation near the condensation threshold, where the nonlinearities in Eq. \eqref{eq:FullModel} can be treated perturbatively by assuming the condensate wave function $\Psi = (\psi_+ e^{il\varphi} + \psi_- e^{-il\varphi})\Psi_l$ with $l\in \mathbb{N}$ the angular number and $\Psi_l(r)$ the radial part of the linearized trapped mode, at which the condensation occurs \cite{Chestnov2024}.
%We then introduce the classical pseudospin $\mathbf{S} = b \psi^\dagger \boldsymbol{\sigma} \psi/\gamma$ with $\psi = [\psi_+,\psi_-]^T$, $\boldsymbol{\sigma}$ the Pauli vector, and the reservoir-condensate scattering efficiency $b = 2\pi \beta \int_0^\infty |\Psi_l|^2 n_r r dr/\hbar$.
Under proper normalization using the characteristic scales of time $\gamma^{-1}$ and space $\sqrt{\hbar/(m\Gamma)}$, the superfluid rotation is described with the pseudovector $\mathbf{S} = \psi^\dagger \boldsymbol{\sigma} \psi$ with $\psi = [\psi_+,\psi_-]^\text{T}$ and $\boldsymbol{\sigma}$ the Pauli vector.
Keeping the three allowed azimuthal harmonics $N_0$, $N_1$, $N_2$ of the dimensionless reservoir density angular dependence $N(\varphi) = N_\text{th} + N_0 + N_1 \cos(2l\varphi) + N_2 \sin(2l\varphi)$ with the threshold value $N_\text{th} = \Gamma/\gamma$, we approximate the system dynamics with coupled vector equations:
\begin{subequations}\label{eq:spindyn}
    \begin{align}
        \label{eq:dS}
        \dot{\mathbf{S}} = N_0 \mathbf{S} + S\mathbf{N} - \left[ (\varepsilon\mathbf{N}+\xi\mathbf{S}) \times \mathbf{S}_\perp \right],
        \\
        \label{eq:dnii}
        \dot{\mathbf{N}} = -(1+2S)\mathbf{N} - (N_\text{th}+N_0)\mathbf{S}_\parallel,
        %\\
        %\label{eq:dn0}
        %\dot{N_0} = W - (1+2S)N_0 -(SN_\text{th}+\boldsymbol{S}\cdot\boldsymbol{N}_\parallel),
    \end{align}
\end{subequations}
where $\mathbf{S}_\parallel$ and $\mathbf{S}_\perp$ are the superfluid vector projections onto the $xy$ plane and the $z$ axis respectively, $\mathbf{N} = (N_1,N_2,0)$ is the normal component vector, and $\xi$, $\varepsilon$ determine the strength of condensate self-interaction and the condensate-reservoir interaction respectively.
The normal component vector evolution \eqref{eq:dnii} is coupled to the equation for the azimuthally uniform exciton density offset:
\begin{equation} \label{eq:dn0}
    \dot{N_0} = W - (1+2S)N_0 -(SN_\text{th}+\mathbf{S}\cdot\mathbf{N}),
\end{equation}
where $W$ is the pumping power excess over the threshold value.

The system of Eqs. (\ref{eq:spindyn},\ref{eq:dn0}) supports a pair of stationary superfluid solutions $S_z=\pm W/(2N_\text{th})$, $\mathbf{S}_\parallel=0$ $\mathbf{N}=0$, shown with blue dots in Fig \ref{Fig1}c, and a continuum of azimuthal standing wave solutions with $S_z = 0$ and $\mathbf{N} = -2N_0\mathbf{S}/S$, shown with red circles.
The spatial density profiles, as well as the condensate phase distributions, corresponding to one of the superfluid states (1), and to a supersolid phase (2) selected from the continuum and marked with red dots in Fig. \ref{Fig1}c, are shown in Fig. \ref{Fig1}d.
The two types of states are characterized by partially overlapping regions of stability in the parameter space ($\varepsilon$, $\xi$, $N_\text{th}$, $W$), and the supersolid phase is reliably reached in the case of strong repulsive condensate-reservoir interactions $\varepsilon>1$ in a wide range of parameters, where the superfluid phase is unstable \cite{Chestnov2024}.

\begin{figure}
    \includegraphics[width=\linewidth]{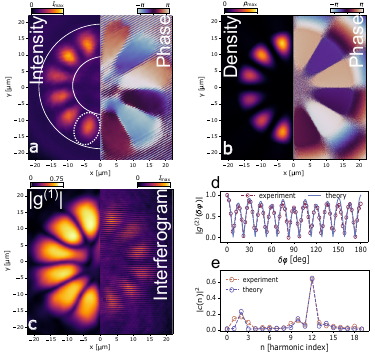}
     \caption{
     Polariton supersolid phase forming in a radially symmetric annular optically induced trap.
     \textbf{a}, Experimentally measured and \textbf{b}, analytically calculated intensity (left) and phase (right).
     \textbf{c}, The spatial map of $|g^{(1)}(\delta \mathbf{r})|$ correlation function (left), extracted from the measured interference pattern (right), revealing uniform spatial coherence of the supersolid state.
     \textbf{d}, The angular dependence of the density correlation function $|g^{(2)}(\delta\varphi)|$, computed within the region enclosed by solid circles in panel a).
     \textbf{e}, The angular density contributions $|c(\mathrm{n})|^2$ of angular density harmonics of index $\mathrm{n}$, showing weak coupling of adjacent harmonics $l=\pm5,7$ to the condensate modes $l=\pm6$.
     }
    \label{Fig2}
\end{figure}  

The results of our experimental study of the supersolid phase are summarized in Fig. \ref{Fig2}.
An example of spatial distribution of the coherent polariton emission intensity and its phase, extracted using interferometry measurements (see Methods), and the corresponding profiles obtained with the semianalytic two-mode model, are shown in Fig.\ref{Fig2}a and Fig. \ref{Fig2}b respectively.
The coherence of the condensate is characterized with the first-order two-point correlation function 
%$g^{(1)}(\mathbf{r},\mathbf{r}^\prime) = \langle \Psi(\mathbf{r}) \Psi(\mathbf{r^\prime}) \rangle$
$g^{(1)}(\delta\mathbf{r}) = \langle \Psi(\mathbf{r}) \Psi(\mathbf{r}+\delta\mathbf{r}) \rangle$, obtained from the interference of the condensate emission with the reference signal at the point $\mathbf{r}$ (see Methods).
The off-diagonal long-range order is demonstrated with the nonvanishing emission coherence, presented in Fig. \ref{Fig2}c.
The crystalline order is, in turn, characterized with the two-point density correlation function $g^{(2)}(\delta\varphi) = \langle |\Psi(\varphi)|^2 |\Psi(\varphi+\delta\varphi)|^2 \rangle_\varphi$, shown in Fig. \ref{Fig2}d, with the angular condensate $|\Psi(\varphi)|^2$ density, obtained by integrating the emission intensity on the annular disk, shown in Figs.\ref{Fig2}a,b.
The angular harmonic decomposition of the condensate density, shown in \ref{Fig2}e, indicates weak coupling of the main state, formed by modes with angular momentum $l=\pm 6$ with adjacent modes $l=\pm 5,7$.
This coupling, as well as the angular dependence of the condensate density and phase and thus the density correlation, is reproduced by accounting for the weak potential slope, stemming form the wedged cavity geometry \cite{Voronova2025}, with the first-order perturbation theory (see Supplementary Materials).

\section{Nambu-Goldstone modes}
\begin{figure*}[t]
    \includegraphics[width=1.0\linewidth]{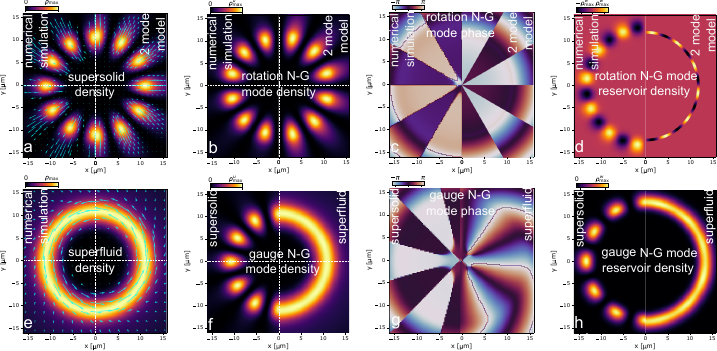}
    \caption{Spatial structure of numerically computed zero-energy N-G modes.
    \textbf{a}, The excitation, associated with spontaneous spatial rotation of the supersolid state, is represented with its spatial density \textbf{b}, $|u|^2$, \textbf{c}, phase $\text{arg}(u)$ and \textbf{d}, normal component perturbation $w$.
    The supersolid condensate and the rotational N-G mode (\textbf{a}-\textbf{d}) are obtained with numerical simulation of the full model \eqref{eq:FullModel} (left) and with approximate 2-mode model \eqref{eq:spindyn} (right).
    \textbf{e}, The persistent superfluid polariton currents, illustrating spontaneous phase winding in the superfluid phase, are shown with cyan arrows.
    \textbf{e}-\textbf{h}, the gauge N-G modes, associated to the global phase rotation of the supersolid (left) and the superfluid (right) state.
    }
    %Panel a illustrate the supersolid state particle density, characterized by the presence of two Nambu–Goldstone (N-G) modes: one associated with rotational symmetry breaking (panels b–d, left and right sides) and another corresponding to global phase symmetry breaking (panels f–h, left side). In contrast, the superfluid state (panels e, f–h at the right side) exhibits only a single N-G mode linked to global phase symmetry breaking. Each column presents: particle density $|\Psi_0|^2$ (first column), density of the perturbation term $|u|^2$ (second column), phase of the perturbation term $\arg(u)$ (third column), and reservoir excitation density $w$ (fourth column). For the supersolid phase, the rotational N-G mode is also obtained using a two-mode approximation \eqref{eq:Jacobian} (panels a–d, right side) in addition to the numerical subspace eigenproblem solution \eqref{eq:BdGnumerical} (see Methods). The arrows highlight differences in the probability current between the supersolid and superfluid states.}
    \label{fig:Goldstone}   
\end{figure*}

The collective excitation spectra of optically trapped polariton condensates are size-quantized and can be obtained numerically following Ref. \cite{Wouters2007a} in the cases of stable superfluid and supersolid phases, shown in Fig. \ref{fig:Goldstone}a and Fig. \ref{fig:Goldstone}e respectively (see details in Methods and Supplementary Materials).
The gapless N-G modes are present in both cases at zero energy, however, while the superfluid is characterized by only one such mode due to spontaneously broken $U(1)$ gauge symmetry, the supersolid phase has an additional zero-energy mode, stemming from the broken radial symmetry.
The spatial structures of zero-energy N-G modes, describing collective excitations $\Psi = \Psi_0 + u + v^*$, $n = n_0 + w + w^*$ over stable states $\Psi_0$, $n_0$, are shown in Figs. \ref{fig:Goldstone}b-d,f-h.
Figs. \ref{fig:Goldstone}b-d show the zero-energy N-G modes, originating from the broken continuous radial symmetry, only present in the supersolid phase, while Figs. \ref{fig:Goldstone}f-h compare the supersolid and the superfluid zero-energy N-G modes, stemming from the spontaneously broken gauge symmetry.

Notably, the rotational N-G mode emerging in the supersolid phase can also be reproduced analytically within the simplified two-mode model.
The zero eigenvalue of the Jacobian matrix of Eqs. \eqref{eq:dS} for the continuum of supersolid states with $S_z=0$ and $|\mathbf{S}_\parallel|=S$ corresponds to coupled in-plane perturbations $\delta\mathbf{N} \propto \delta \mathbf{S}_\parallel \perp \mathbf{S}_\parallel$.
The corresponding spatial profiles of the underlying N-G mode, shown at the right-hand sides of Fig.\ref{fig:Goldstone}b-d, qualitatively and symmetrically reproduce their numerically computed counterparts, shown at the respective left-hand parts.

\section{Discussion}

We demonstrated the formation of a supersolid polariton condensate phase, characterized with diagonal and off-diagonal long-range order, in axially symmetric optically induced traps, as summarized in Fig. \ref{Fig2}.
The combination of the condensate density spatial distribution measurements with the interferometry technique employed for extracting the first-order correlation function reveals pronounced azimuthal modulation of both values sharing the same crystalline ordering.
At the same time, observation of global spatial coherence of the condensate emission reveals long-range off-diagonal order in the condensate and allows describing it with a coherent order parameter within the mean-field approximation \eqref{eq:FullModel}.

The coexisting long-range off-diagonal and crystalline order in the supersolid phase is reproduced by accounting for polariton short-range interactions and the coupling between the condensate and the normal component.
We also stress the insufficiency of the widely used simplified nonlinear model, where the equation for the normal component density is adiabatically eliminated, failing to describe stable polariton states with spontaneously broken azimuthal symmetry in the case of axially symmetric traps.
The full nonadiabatic model also predicts supersolid phase instability in the case of weak condensate-reservoir interactions, highlighting the significance of emergent reservoir-mediated attraction for stability of the supersolid phase \cite{Chestnov2024}.

The physical mechanism, simultaneously destabilizing the uniform superfluid and stabilizing the supersolid phase, shares similarity with the condensate auto-localization through reservoir depletion \cite{Estrecho2018}.
In this scenario, the condensate spontaneously and locally burns holes in the reservoir density due to avalanche bosonic-stimulated scattering and is trapped in the thus-produced potential wells.
In the case of uniform pumping, such autolocalized condensates are stochastically positioned in the cavity plane and remain decoupled rather than forming a synchronous coherent state.
However, in the case of annular pumping, the competition of potential trapping due to the auto-localization mechanism with the rotational motion induced by the dissipative mode selection, prevents spatial separation of the condensate into decoupled droplets.
This interplay results in the formation of the coherent supersolid phase that is robust to deformation and disorder.
The lattice constant $a$ of the supersolid phase is governed by the strength of interactions and the scaling $a\propto1/\sqrt{R}$ in the limit of large trap radii $R$ (see Supplementary Materials).

Within the mean-field two-component model, reproducing our experimental observations, we further illustrate the supersolidity with the emergence of zero-energy N-G modes, specific to spontaneously broken continuous spatial symmetry, as shown in Fig. \ref{fig:Goldstone}.
The comparison between the N-G modes of the supersolid phase and that of the superfluid state
%, stable in the range of parameters beyond the experimental reach,
vividly demonstrates the hallmark supersolid properties of the collective excitation spectrum.
In contrast to polariton supersolidity in bound-in-continuum polariton condensates \cite{Trypogeorgos2025}, our setup does not require an explicit discrete translational symmetry of a photonic lattice.
Moreover, in our implementation of nonequilibrium supersolidity, the coherent polariton condensate is wholly in the supersolid phase rather than being superimposed with a stronger populated spatially uniform superfluid fraction.
This renders experimental probing of collective excitations demonstrating gapless N-G modes of the supersolid phase, as well as other aspects of polariton supersolidity in our system, a promising direction of further studies.

\section{Methods}

\bmhead{Sample}
The sample is a high-homogeneity, ultrahigh-finesse planar microcavity ($Q > 10^4$, polariton lifetime $\tau_p$ is on the order of tens of picoseconds) grown by molecular-beam epitaxy on a GaAs substrate. It consists of a $5\lambda/2$ Al$_{0.3}$Ga$_{0.7}$As cavity sandwiched between distributed Bragg reflectors, with the top (bottom) comprising 45 (50) pairs of AlAs/Al$_{0.15}$Ga$_{0.85}$As $\lambda/4$ layers. Embedded within the cavity at the anti-nodes of the optical field are four sets of three GaAs quantum wells (each well 12 nm~thick, separated by 9~nm Al$_{0.3}$Ga$_{0.7}$As barriers). The cavity gradient allowed selection of the exciton-photon detuning; experiments were performed at small negative detuning. The sample was cooled to 6~K in a closed-cycle low-vibration cryostat (sample vibration $< 100$~nm) and excited non-resonantly in the normal direction. The optical annular trap was created using a continuous-wave Ti:sapphire laser tuned to 762~nm (corresponding to a reflectivity dip next to the stop band) and a digital micro-mirror device based spatial light modulator. The emission of condensate (810~nm) was collected normal to the surface and isolated from scattered pump light using a long-pass filter (800~nm).

\bmhead{Interferometric measurements}
%A time-resolved Mach-Zehnder interferometer was implemented for detection. 
The reference arm of a time-resolved Mach-Zehnder interferometer incorporated a telescopic lens system (focal lengths selected for 10$\times$ magnification) to spatially expand emission from a single condensate petal (highlighted with a dotted line in Fig.~\ref{Fig2}a), while the signal arm transmitted the full condensate distribution. A diaphragm spatially filtered the magnified reference-arm emission, generating residual Airy fringes (Fig.~\ref{Fig2}c). The output optical cube facilitated overlap tuning between arms, and lateral/vertical retroreflector adjustments enabled controlled wavevector mismatch to set interference fringes slope. A piezo-mounted retroreflector in the signal arm served as a delay line. Interference patterns (Fig.~\ref{Fig2}c) at phase differences of 0, $\pi/2$, $\pi$ and $3\pi/2$ were captured on a CCD after path-length balancing and synchronization of arms.
%\bmhead{Phase and coherence extraction}
The relative phase distribution was extracted using a four-step phase-reconstruction algorithm as in Ref. \cite{Barrat2024}.
The absolute phase distribution was then determined by subtracting the tilt phase component (corresponding to the slope of the interference fringes) from the relative phase.
The procedure for the measurement of coherence is analogous to the one in Ref. \cite{Trypogeorgos2025}, where the two-point correlation function $|g^{(1)}(\mathbf{r},\mathbf{r}+\delta\mathbf{r})|$ is extracted from upper and lower envelopes of the interferogram.

%LOADING... ATTENTION! $g^{(1)}(\mathbf{r},\mathbf{r}^\prime) = \langle \Psi(\mathbf{r}) \Psi(\mathbf{r^\prime}) \rangle$
%The two-point correlator is proportional to the visibility $V(\mathbf{r}) = (I_u(\mathbf{r})-I_l(\mathbf{r}))/(I_u(\mathbf{r})+I_l(\mathbf{r}))$ of the interferometric fringes, where $I_u(\mathbf{r})$ and $I_l(\mathbf{r})$ is the upper and lower envelope of the interferogram. We calculate $V$ by first extracting the $\mathbf{k} = \mathbf{k}_\mathbf{r}$ and $\mathbf{k} = 0$ components of the interferogram by appropriate filtering in the reciprocal space and then using the Hilbert transform to calculate the envelope of the $\mathbf{k}_\mathbf{r}$ component; the visibility is then simply the envelope over the $\mathbf{k} = 0$ component. For imbalance between the two images, $V$ needs to be rescaled by $2\sqrt{I_sI_{ref}}/(I_s + I_{ref})$, where $I_{s,ref}$ are the intensities in signal and reference arm of the interferometer respectively, to give a correctly normalised $g^{(1)} (\mathbf{r},\mathbf{r}')$. However, this normalization does not fully compensate for inhomogenity of $I_{ref}$ in the overlap of the two images distorts the density modulation.

\bmhead{Numerical simulations}
The equations~\eqref{eq:FullModel} were solved numerically using a GPU-accelerated Split-Step Fourier method. Periodic boundary conditions were applied, with the computational domain chosen large enough to avoid boundary effects.
Initial conditions consisted of low-amplitude complex white noise for the condensate, representing spontaneous nature of condensation, and zero for the reservoir.
The pump profile was ring-shaped, given by $P~=~P_0 \exp(-[\sqrt{X^2 + Y^2} - w_1]^2/w_2^2)$, where $P_0$ is the pump power, $w_1$ the trap radius, and $w_2$ the trap width.
The condensate evolution was simulated over the time interval long enough to reach a stable single-mode solution.
The subspace eigenproblem for the sparse matrix~\eqref{eq:BdGnumerical}, based on the precomputed stationary solution of~\eqref{eq:FullModel}, was solved to identify Nambu--Goldstone modes.
The shift-invert Arnoldi algorithm was employed to compute the eigenmodes with the smallest eigenvalue absolute values. Parameters used in simulations: \cite{nparams}.

%\bmhead{Acknowledgements}
%The authors express their gratitude to B. Altshuler and G. Shlyapnikov for fruitful discussions.
%A.N. acknowledges support by the Russian Science Foundation under Grant No. 25-12-00135.
%A.L., R.C., M.C., and A.V.K. acknowledge support from St. Petersburg State University (Grant No. 125022803069-4).
\bmhead{Competing interests} The authors declare no competing interests.
\bmhead{Data availability} The data of this study is available from the corresponding author upon reasonable request.
\bmhead{Author contributions}
P.K. performed numerical simulations and explained the theoretical concepts to A.L. and M.C.
I.C. contributed to interpretation and representation of analytical and experimental results.
R.C. supervised experimental work, performed by A.L. and M.C.
A.N. and A.K. supervised the work, A.N. proposed the idea of experiment and its interpretation.
All authors contributed to discussions and editing of the manuscript. 

\bibliography{references.bib}

\onecolumn
\include{supplementary}
\end{document}

%% file: supplementary.tex
\begin{center}
  \textbf{\large Supplementary Information}
\end{center}

\setcounter{equation}{0}
\setcounter{figure}{0}
\setcounter{table}{0}
\setcounter{section}{0}
\setcounter{page}{1}
\renewcommand{\theequation}{S\arabic{equation}} 
\renewcommand{\thefigure}{S\arabic{figure}}

\section{Experimental setup}

\begin{figure}[h!]
    \centering
    \includegraphics[width=0.7\linewidth]{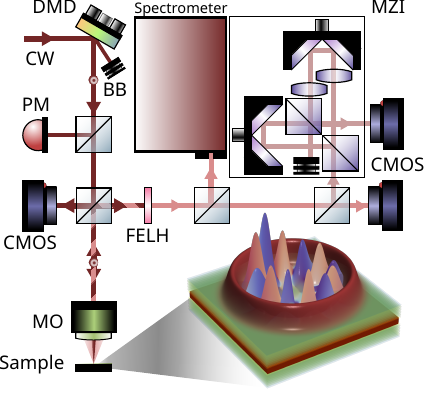}
    \caption{Schematics. DMD — spatial light modulator based on digital micromirror device. CW — single-mode cw 762 nm laser. BB — black body. PM — power meter. MO — micro objective. FELH — low-pass spectral filter. MZI — Mach-Zhender interferometer. CMOS — a digital camera with an active pixel sensor.}
    \label{figS1}
\end{figure}

Optical pumping was achieved using a continuous-wave (CW) single-mode diode laser (Toptica TA Pro) with the wavelength tuned to the wavelength 762~nm at the first reflectivity dip of the top distributed Bragg reflector.
The laser beam passed through an acousto-optic modulator (AOM, not shown in the Fig. \ref{figS1}) for power control (range: 0–30~mW) and a microelectromechanical based (digital microelectromechanical device, DMD) spatial light modulator (SLM, Texas Instruments DLP6500, pixel pitch 7.56~µm, $>2$M~mirrors) to generate a ring-shaped optical trap.
This trap, defined by the SLM pattern, featured an inner diameter of $\sim30$~µm and a width of $\sim1$~µm, optimized to populate the state of exciton-polaritons with orbital angular momentum (OAM) $|L|=6$.
The beam was focused onto the sample surface via a 4-mm focal length Mitutoyo M Plan APO NIR 50x objective (NA = 0.42) in reflection geometry, with demagnification controlled by a 150-mm focal length lens (not shown in the Fig. \ref{figS1}, demagnification factor 4:150).
The pump power near the lasing threshold $\sim15$~mW was maintained using a half-wave plate and a linear polarizer, with time modulation by AOM mitigating sample heating.

\newpage

\begin{figure}[h!]
    \centering
    \includegraphics[width=1.0\linewidth]{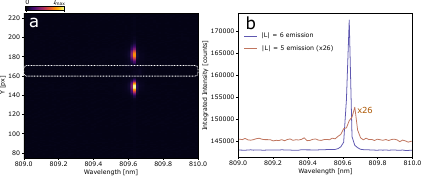}
    \caption{The energy-resolved emission of the polariton condensate confined in the ring-shaped trap. a) The part of the emission cut out by the spectrometer slit refocused onto the CCD. b) The integrated emission spectrum from the entire matrix and from the matrix region highlighted by dotted lines to demonstrate the resolution of neighboring states with OAM $|L|=6$ and $|L|=5$ (amplified by 26 times), respectively.}
    \label{figS2}
\end{figure}

Polariton emission was collected in the reflection geometry through the same 50x objective.
Backscattered pump light was suppressed using a long-pass spectral filter (FELH, cutoff 800 nm). The PL was then split into two analysis pathways:

1. Spectroscopy Path. The condensate emission directed to the back focal plane of the objective was coupled into a 0.5-m imaging spectrometer (Acton equipped with a PIXIS-256 CCD camera).
The optical system and pump power were optimized to generate a spectrally pure condensate state, verified by the absence of emision at the trap center from the state with OAM $|L|=6$.
The spectrometer’s resolution of 15~µeV resolved neighboring states.
Emission at the trap center was attributed solely to the condensate density at the weakly coupled state with OAM $|L|=5$ due to spectral separation. 

2. Interferometry Path. The emission routed to the customized Mach-Zehnder interferometer, where the signal arm transmitted the full emission and contained a dual-stage delay line combining precision mechanical translation and piezo actuation for path-length control. The reference arm employed a telescopic system with two lenses ($f_1 = 25$~mm, $f_2 = 250$~mm, providing 10x magnification) to spatially expand a single condensate petal into a reference plane wave. The interferograms were recorded using a thermoelectrically cooled camera (MAX04AM, GSENSE2020e NIR CMOS sensor, exposure time about 0.1 s).

\newpage

\section{Deformation in a uniform gradient potential}

A weak gradient potential $ax = ar\cos(\varphi)$ perturbatively couples vortex states $L$ with adjacent states with adjacent angular momenta.
Namely, the first order perturbation theory yields coupling with the modes $l\pm 1$.
The perturbed vortex state reads
\begin{equation} \label{eq:Psidef}
    \Psi_l^{(1)}(r,\varphi) = c_l^{(1)} \Psi_l^{(0)} + c_{l+1}^{(1)} \Psi_{l+1}^{(0)} + c_{l-1}^{(1)} \Psi_{l-1}^{(0)},
\end{equation}
where the first order coefficients are given by
\begin{equation}
    c_{l\pm1}^{(1)} = {a\pi } { \int_0^\infty R^*_{l}(r) R_{l\pm1}(r) r^2 dr   \over E_l - E_{l\pm 1}}.
\end{equation}
Neglecting the difference in the radial parts of the adjacent modes $R_l(r) \approx R_{l\pm1}(r)$ of the involved azimuthal modes and assuming the same distances $|\Delta|$ to the adjacent states $E_{l\pm1}=E_l \pm \mathrm{Re}\lbrace \Delta \rbrace - i \mathrm{Im}\lbrace \Delta \rbrace$ on the complex energy plane for simplicity, one has
\begin{equation}
    c_{l\pm1}^{(1)} \approx {a R \over 2|\Delta|^2} ( \mp \mathrm{Re}\lbrace \Delta \rbrace - i \mathrm{Im}\lbrace \Delta \rbrace).
\end{equation}

\begin{figure}[h!]
    \centering
    \includegraphics[width=\linewidth]{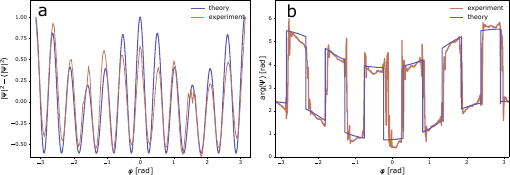}
    \caption{Angular modulation of the condensate density (a) and phase (b). Experimental data, corresponding to Fig. \ref{Fig2}, is shown with blue lines, and compared with the results of the first-order perturbation theory, shown with red lines.}
    \label{figS3}
\end{figure}

The couple of degenerate deformed vortex states \eqref{eq:Psidef} with the opposite rotation directions forms the basis for the weakly nonlinear model \eqref{eq:spindyn} and, similarly to the isotropic case, yields stable condensate solutions of the form $\Psi_l^{(1)}+e^{i\theta}\Psi_{-l}^{(1)}$, conserving the chiral symmetry.
An example of such a state, shown in Fig. \ref{Fig2}b for $\theta = 0$ and $aR\Delta/2|\Delta|^2 = 0.5i$, is produced using the expression for the polariton condensate wavefuction \cite{Chestnov2024}:
\begin{equation}
    \Psi_l^{(0)} (r,\varphi) = \exp(il\varphi)\times \left\lbrace \begin{matrix}
        AJ_l(\varkappa r), & r<R \\
        BH^{(1)}_l(\varkappa r), & r>R
    \end{matrix} \right.,
\end{equation}
with $\varkappa = \sqrt{2m(\omega_0 + i \Gamma/2)/\hbar}$ and $\omega_0$ the condensate emission frequency at the condensation threshold.
It is in qualitative agreement with experimental data, as shown in Fig. \ref{Fig2}d,e and Fig.\ref{figS3}.

\section{Collective excitations}

Eqs. \eqref{eq:FullModel} can be linearized near fixed point solutions $\Psi = \Psi_0e^{-i\omega_0t}$, $n=n_0$ by assuming small perturbations:
\begin{subequations}\label{eq:perturb}
    \begin{eqnarray}
        \label{eq:pertpsi}
        \Psi = e^{-i\omega_0t}\left[\Psi_0 + ue^{-i\omega t} + v^*e^{i\omega^* t}\right]
        \\
        \label{eq:pertn}
        n = n_0 + w e^{-i\omega t} + w^* e^{i\omega^* t}
    \end{eqnarray}
\end{subequations}

The collective excitation spectrum is given by the matrix equation $\hbar\omega[u,v,w]^\text{T} = L[u,v,w]^\text{T}$ with

\begin{equation}\label{eq:BdGnumerical}
    L = \left[ \begin{matrix}
        \hat T + {\alpha+i\beta \over 2} n_0  - i{\hbar \Gamma \over 2} + 2 \alpha_1 |\Psi_0|^2 - \hbar \omega_0 & \alpha_1 \Psi_0^2 & {\alpha+i\beta \over 2} \Psi_0 \\
        - \alpha_1 (\Psi_0^*)^2 & -\hat T + {-\alpha+i\beta \over 2} n_0  - i{\hbar \Gamma \over 2} - 2 \alpha_1 |\Psi_0|^2 + \hbar \omega_0 & {-\alpha+i\beta \over 2} (\Psi_0)^* \\
        -i \beta n_0 (\Psi_0)^* & -i \beta n_0 \Psi_0 & - i\hbar (\gamma+{\beta \over \hbar}|\Psi_0|^2) & 
    \end{matrix} \right]
\end{equation}

where $\hat{T} = -{\hbar \nabla^2}/2m$ is the kinetic energy operator.

The excitations in the two-mode model can be obtained from the Jacobian of the system \eqref{eq:FullModel}:
\begin{equation}\label{eq:Jacobian}
    J = \left[ \begin{matrix}
        N_0 +{N_1\over2}{S_x\over S} & {N_1\over2}{S_y\over S}+\xi S_z & {N_1\over2}{S_z\over S} + \varepsilon{N_2\over2}+\xi S_y & S_x & S\over2 & \varepsilon{S_z\over2}\\
        {N_2\over2}{S_x\over S}-\xi S_z&N_0+{N_2\over2}{S_y\over S}&{N_2\over2}{S_z\over S}-\varepsilon{N_1\over2}-\xi S_x & S_y &-\varepsilon{S_z\over2} & S\over2 \\
        -\varepsilon{N_2\over2} & \varepsilon{N_1\over2} & N_0 & S_z & \varepsilon {S_y\over2} & -\varepsilon{S_x\over2}\\
        -2(N_\text{th}+N_0){S_x\over S}-N_1 & -2(N_\text{th}+N_0){S_y\over S}-N_2 & -2(N_\text{th}+N_0){S_z\over S} & -(1+2S) & -S_x & -S_y \\
        -2N_1{S_x\over S}-2(N_\text{th}+N_0) & -2N_1{S_y\over S} & -2N_1 {S_z\over S} & -2S_x & -(1+2S) & 0 \\
        -2N_2{S_x\over S} & -2N_2{S_y\over S}-2(N_\text{th}+N_0) & -2N_2{S_z\over S} & -2S_y & 0 & -(1+2S)        
    \end{matrix} \right]
\end{equation}

\begin{figure}[h!]
    \centering
    \includegraphics[width=\linewidth]{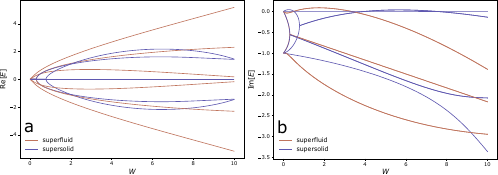}
    \caption{Real (a) and imaginary (b) parts of the elementary excitations spectra obtained within the two-mode approximation as the eigenvalues of the Jacobian matrix \eqref{eq:Jacobian} at fixed points, corresponding to superfluid vortices (blue) and supersolid states (red), as functions of the pumping power $W$.
    The presence of exactly zero-energy Nambu-Goldstone mode in the supersolid case illustrates spontaneous breaking of the continuous radial symmetry.
    Parameters: $\varepsilon = \xi = N_\text{th} = 3$.}
    \label{figS4}
\end{figure}

For the stable supersolid state, by aligning the axes with the wavefunction antinodes, one has $N_2=S_y=0$ and the Jacobian takes the block structure
\begin{equation}
    J = \left[ \begin{matrix}
        J_1 & 0 \\
        0 & J_2
    \end{matrix}\right],
\end{equation}
with the blocks given by
\begin{subequations}
\begin{align}
    J_1 =  \left[ \begin{matrix}
        0 & S & S/2 \\
        -2N_\text{th} & -(1+2S) & -S \\
        2(N_0-N_\text{th}) & -2S & -(1+2S)
    \end{matrix}\right], \\
    J_2 =  \left[ \begin{matrix}
        N_0 & \varepsilon N_0 - \xi S & S/2 \\
        -\varepsilon N_0 & N_0 & -\varepsilon S/2 \\
        -2(N_\text{th}+N_0) & 0 & -(1+2S)
    \end{matrix}\right],
\end{align}
\end{subequations}
where the azimuthally uniform normal density harmonic $N_0$ reads \cite{Chestnov2024}
\begin{equation}
    N_0 = {1 \over 2}\left[3 N_\mathrm{th} + W - \sqrt{(3N_\mathrm{th}+W)^2 - 4N_\mathrm{th}W} \right].
\end{equation}
The eigenvalue equation for the second block $J_2$ has a zero free term
\begin{equation}
    \left( \varepsilon \xi S - \varepsilon^2N_0 - 1 \right) \left[ S(N_0-N_\text{th}) +N_0 \right] = 0,
\end{equation}
The eigenvector, corresponding to the zero-energy Nambu-Goldstone boson, has only two nonzero components $\delta S_y$ and $\delta N_2 = -2(N_\text{th}+N_0)/(1+2S)\delta S_y$.
Complex excitation spectra, computed as eigenvalues of the Jacobian matrix \eqref{eq:Jacobian}, illustrating the difference between the superfluid vortex and supersolid states, are presented in Fig.\ref{figS4}.
Note that the gapless Nambu-Goldstone mode, stemming from spontaneous breaking of gauge symmetry, responsible for the formation of the condensate phase, is absent in the analysis based on the system of Eqs. \eqref{eq:spindyn}, where the phase dynamics is integrated out.

\section{Supersolid period scaling}
\begin{figure}[h!]
    \centering
    \includegraphics[width=1.0\linewidth]{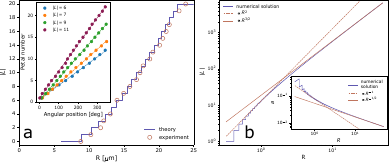}
    \caption{
\textbf{a}, The condensate angular index absolute value $|L|$ as a function of the trap radius $R$.
Experimental data closely follows the step-like theoretical prediction, confirming the incremental growth of the condensate angular index $|L|$ with increasing confinement size.
The dimensions are rescaled for clarity, with characteristic length $R_0 = 15\, \mu$m and offset $4\, \mu$m, corresponding to the radial broadening of the normal component ring. 
\textit{Inset:} The azimuthal positions of density modulation lobes for four representative experimental datasets ($|L| = 6, 7, 9, 11$); the linear dependence indicates the emergence of crystalline density order in larger traps.
(b) The log-log plot of the numerically evaluated dependence $|L|(R)$.
Reference lines indicate the large-$R$ asymptotic limits: $|L| \propto R^{2}$ (dashed red), as reported in \cite{Dreismann2014b}, and $|L| \propto R^{3/2}$ (solid red), accounting for the mode competition.
Inset: radial dependence of the lattice constant $a(R)$: our result, $a \propto R^{-1/2}$ (solid red), compared to the prediction from \cite{Dreismann2014b}, $a \propto R^{-1}$ (dashed red).
All quantities are shown in the units of $R_0$.
Interaction parameter: $\varepsilon = 3$.}
    \label{figS5}
\end{figure}

The formation of stable polariton supersolid states spontaneously breaking the continuous spatial symmetry is explained in the nonlinear model \eqref{eq:FullModel}.
However, the selection of the orbital angular momentum $l$ is governed by the dissipative mode competition mechanism, which is captured in the linear model valid near the condensation threshold.
To be specific, Eqs. \eqref{eq:FullModel} with omitted nonlinear terms in $\Psi$ and stationary normal component density $n$, inheriting the spatial shape from that of the pumping, reduce to a non-Hermitian eigenvalue problem.
The condensation occurs at a certain pumping power, at which one of the polariton modes, characterized with radial and angular numbers $\mathrm{N}$ and $l$ and the energy $E_{\mathrm{N},l}$, reaches the exponential growth regime $\mathrm{Im}\lbrace E_{\mathrm{N},l} \rbrace > 0$.

Solving $\mathrm{Im}\lbrace E_{\mathrm{N},l} \rbrace = 0$ for mode-specific critical pumping powers $P_{\mathrm{N},l}$ and minimizing among all modes yields the condensation threshold power and specifies the mode at which it occurs.
We used a semi-analytical approach, based on matching piecewise-defined wave functions under the approximation of narrow pumped region $P=P_0\delta(r-R)/R$.
This approach shows that the angular mode index of the condensate $l$ exhibits superlinear monotonous growth with the trap radius $R$, while the radial index remains at $\mathrm{N} = 0$, which is in agreemennt with our observations.
The dependence $l(R)$, governed by the parameters $\varepsilon = \alpha/\beta$ and $R_0 = \sqrt{\hbar/(m\Gamma)}$ was found to be in quantitative agreement with our experimental observations, as shown in Fig. \ref{figS5}.

In Ref. \cite{Dreismann2014b} this superlinear growth, observed in a similar configuration, was explained using the asymptotic analysis of the stable solutions of complex Ginzburg-Landau equation neglecting the mode competition.
This analysis yielded the quadratic dependence $l\propto R^2$, which is in good qualitative agreement with both the experimental observations and with our semi-analytical approach in the intermediate range of $l\sim10$, as shown in Fig. \ref{figS5}b.
In contrast, our approach reveals steeper growth in the case of smaller traps ($l<10$) and the asymptotic behaviour $l\propto R^{3/2}$ in the limit of large trap radii ($l>100$).

The scaling $l\propto R^{3/2}$ activates in the limit of large traps, where the linear threshold pumping density saturates at a certain level as the condensate becomes effectively one-dimensional.
This size-independent pumping power induces an annular potential barrier of a fixed height, compensating the centrifugal force $F_c$ for the rotating polariton fluid.
Approximating the rotation speed as $v = \hbar l/(mR)$ and using the expression for the centrifugal force $F_c = m v^2/R$, we transform the condition $F_c(R) = \mathrm{const}$ to $l^2/R^3 = \mathrm{const}$, which results in the scaling law $l\propto R^{3/2}$.

This dependence of the angular number $l$ corresponds to the scaling of the supersolid lattice constant $a \propto 1/\sqrt{R}$, which is slower than the earlier predicted inverse linear scaling  of the lobe separation distance $a \propto 1/R$, as shown in Fig.\ref{figS5}b.
Weak dependence of the supersolid period on the trap size allows a wide range of trap sizes where it can be experimentally observed with standard optical tools, constrained by the wave limit condition $a > \lambda$ with $\lambda\sim1\,\mu$m the wavelength of polariton emission.
The supersolid period is also limited from above by the minimal trap size, at which the condensation occurs at a degenerate doublet of excited counter-rotating modes.
In the case of sufficiently strong interactions $\varepsilon>1$, required for supersolid phase formation, the upper limit for the supersolid period is given by $R_0\sim10\,\mu$m.
Within the allowed range $a<R_0$, the supersolid period can take any value depending on the trap size, imposing periodic boundary conditions, and the strength of interactions.

%% file: main.bbl
\begin{thebibliography}{39}
\bibitem[\protect\citeauthoryear{}{}]{nparams}
\begin{botherref}
Set of parameters used for superfluid and supersolid solutions.
Set of parameters used for superfluid solution: $\Gamma = 0.1~$ps$^{-1}$,
  $\alpha_1 = 12.5 ~\mu$eV$\mu$m$^2$, $\alpha = 15.0 ~\mu$eV$\mu$m$^2$,
  $\frac{\beta}{\hbar} = 0.01 ~$ps$^{-1}\mu$m$^2$, $\gamma = 0.05~$ps$^{-1}$,
  $w_1 = 14.0~ \mu$m, $w_2 = 1.6~ \mu$m, $m = 2.0 \cdot10^{-5}m_0$, $P_0 =
  5.25~ $ps$^{-1}\mu$m$^{-2}$; for supersolid solution: $\Gamma =
  0.1~$ps$^{-1}$, $\alpha_1 = 10.0 ~\mu$eV$\mu$m$^2$, $\alpha = 20.0
  ~\mu$eV$\mu$m$^2$, $\frac{\beta}{\hbar} = 0.01 ~$ps$^{-1}\mu$m$^2$, $\gamma =
  0.05~ $ps$^{-1}$, $w_1 = 14.0~ \mu$m, $w_2 = 1.6~ \mu$m, $m = 2.0
  \cdot10^{-5}m_0$, $P_0 = 4.9~ $ps$^{-1}\mu$m$^{-2}$, where $m_0$ is free
  electron mass.
\end{botherref}
\endbibitem
\end{thebibliography}
